\documentclass[lettersize,journal]{IEEEtran}
\usepackage{amsmath,amsfonts}
\usepackage{algorithmic}
\usepackage{algorithm}
\usepackage{array}
\usepackage[caption=false,font=normalsize,labelfont=sf,textfont=sf]{subfig}
\usepackage{textcomp}
\usepackage{stfloats}
\usepackage{url}
\usepackage{verbatim}
\usepackage{graphicx}
\usepackage{cite}
\usepackage{xcolor}
\usepackage{multirow} 
\usepackage{tabularx} 
\usepackage{float} 
\usepackage{threeparttable} 
\usepackage{hyperref} 
\usepackage{mathptmx} 
\usepackage{dcolumn}  
\usepackage{ulem}     
\usepackage{siunitx}
\usepackage[version=4]{mhchem}
\usepackage{braket}
\usepackage{booktabs}
\DeclareMathAlphabet{\pazocal}{OMS}{zplm}{m}{n} 

\hypersetup{
    colorlinks=true,
    linkcolor=blue,
    citecolor=blue,
    filecolor=blue,
    urlcolor=blue
}

\newcommand{\SSD}{Sensor Science Division, National Institute of Standards and Technology, Gaithersburg, Maryland 20899, USA}

\newcommand{\CTL}{Communications Technology Laboratory, National Institute of Standards and Technology, Boulder, Colorado 80305, USA}

\begin{document}

\title{Development of a Quantum Blackbody Thermometer toward Primary On-orbit Thermometry}

\author{Peter~J.~Beierle, Denis~Tremblay, Noah~Schlossberger, Christopher~L.~Holloway, Stephen~P.~Eckel, and Eric~B.~Norrgard
\thanks{Manuscript received XXX XXX, 2026.}
\thanks{\textit{Corresponding author: Eric B. Norrgard}.}%
\thanks{Peter J. Beierle is with the ERT (Greenbelt, Maryland) and the University of Maryland -- College Park Earth Systems Science Interdisciplinary Center, College Park, Maryland  (email: \mbox{peter.beierle@ertcorp.com}).}%
\thanks{Denis Tremblay is with the ERT (email: \mbox{denis.tremblay@ertcorp.com}).}%
\thanks{Noah Schlossberger and Christopher L. Holloway are with the \CTL ~(email: \mbox{noah.schlossberger@nist.gov}; \mbox{christopher.holloway@nist.gov}).}%
\thanks{Stephen P. Eckel and Eric B. Norrgard are with the \SSD~ (email \mbox{stephen.eckel@nist.gov}; \mbox{eric.norrgard@nist.gov}).}
}



\maketitle

\begin{abstract}
{W}{e} present a roadmap to a deployable, intrinsically calibrated thermometer with long-term accuracy of 30 mK, exceeding existing on-orbit resistance-based thermometers. Our quantum blackbody thermometer is based on measuring fluorescence ratios of optically excited rubidium atoms in microfabricated vapor cells. The key advantage of the quantum blackbody thermometer is that long-term stability of the fluorescence ratios is guaranteed by the immutable physical properties (transition strengths) of the rubidium atom. This should be compared against resistance-based thermometers, such as platinum resistance thermometers, which may be calibrated with exceptional accuracy but are susceptible to temporal drift and shifts due to improper handling.
\end{abstract}

\begin{IEEEkeywords}
Calibration, cross-track infrared sounder
(CrIS), infrared observations,  remote sensing, satellite, quantum sensors, temperature sensors
\end{IEEEkeywords}

\section{Introduction}
\IEEEPARstart{F}{o}r Earth-observing radiance instruments (infrared, microwave, etc.), the accuracy of the radiometric product critically depends upon a high-emissivity blackbody: an on-board radiometric calibration source (OBRCS). 
The NASA 2007 decadal survey \cite{NRC2007} recommended a 30 mK radiometric accuracy (unless otherwise stated, all uncertainties are one standard error [$k=1$]). 
Yet, in 2025, inter-satellite comparisons of many infrared sensors still show uncertainty and biases of  roughly 100 mK \cite{Wang2022, Loveless2023}.
It is in general desirable to reduce the overall radiometric uncertainty of such sensors in order to improve the inter-sensor radiometric consistency. Efforts to make up this deficit, including international inter-sensor calibration\cite{goldberg2011global}, are otherwise highly resource intensive. 
Additionally, having a robust on-orbit absolute radiometric reference would greatly enhance radiometric agreement in a long-term climate record, especially in instances if there is a gap in such records. Thus, it is desirable to improve the radiometric uncertainty and biases at their root sources.

The OBRCS temperature is often a leading radiometric uncertainty contributor for sensor data products, as exemplified in the JPSS CrIS \cite{Tobin2013} and JPSS ATMS \cite{Weng2016} series. OBRCS temperatures are typically monitored by platinum resistance thermometers (PRTs) \cite{Strouse2007}, which may differ by as much as 100 mK and may drift by as much as 10 mK/year \cite{Mason1996, Smith2020, Smith2021, Latvakoski2011}. Transport to space may induce additional offsets of up to 250 mK \cite{Tobin2013}. The current state of temperature metrology places a fundamental limitation on the long-term absolute interpretation of geophysical data products.

As two examples, Table \ref{tab:cris_atms} shows the aggregated contributions, or error budget, from different sources of radiometric uncertainty for on-orbit infrared \cite{Tobin2013} and microwave \cite{Weng2016} sounders. For these sensors, temperature knowledge of the OBRCS (referred to as the “internal calibration target” or “ICT” or "warm target") is often the leading contributor that otherwise cannot be reduced after sensor optimization. Thus, improved temperature uncertainty of the OBRCS offers the greatest potential to improve the overall radiometric accuracy of the OBRCS.

To highlight a recent effort to improve on-orbit radiometric uncertainty, CLARREO has an absolute accuracy requirement of 30 mK for a 5 year mission \cite{Wielicki2013}, more stringent than has been realized using PRT sensors. The CLARREO instrument design has two OBRCSs: the first OBRCS is kept at near-constant temperature and is measured by PRTs, while the second OBRCS has a phase change temperature sensor.  The phase change temperature sensor provides precise temperature knowledge, allowing improved radiometric calibration of the first OBRCS. Phase change references can achieve few millikelvin temperature accuracy but are substantially  larger and more massive than PRTs \cite{Topham2015, Peters2024}. And while more stable than PRTs, phase change references provide only a single reference temperature while still being susceptible to errors due to impurities and other environmental factors. 

Here we propose a quantum blackbody thermometer (QBT) as an on board absolute temperature sensor and reference to provide significantly improved accuracy and long term stability to temperature measurement.  A key feature of the QBT is the ability to self-calibrate photodectors used for fluorescence readout.  This results in a primary temperature measurement, in that it is not traceable to a measurement of like kind.  The aim of the QBT is to provide an intrinsically accurate temperature measurement  between 250 K and 400 K. QBT technology also has the potential to replace or complement PRTs, phase change temperature sensors, and other thermometers in a variety of remote sensing, autonomous, or otherwise inaccessible systems \cite{Tew2026}.
We highlight as a potential application an improved radiometric calibration reference with a single OBRCS device monitored by a QBT.

\begin{table*}
\begin{tabularx}{\textwidth}{ Xr||Xr }
 \hline\hline
 \multicolumn{2}{c||}{\textbf{Infrared (CrIS)}} & \multicolumn{2}{c}{\textbf{Microwave (ATMS)}} \\
 Uncertainty Source & Contribution \cite{Tobin2013}& Uncertainty Source & Contribution \cite{Weng2016}\\
 \hline
 Internal Calibration Target(ICT)/ PRT Temperature & 0.033 K & Internal Calibration Target(ICT)/ PRT Temperature & 0.1 K \\
 ICT Emissivity & 0.023 K -- 0.03 K & ICT Emissivity & 0.03 K \\
 Measured ICT Reflection Temperature & 0.01 K& ICT Coupling & $< 0.2$ K \\
 Model ICT Reflection Temperature & 0.007 K -- 0.03 K& Space Target Temperature/ Sidelobe Contamination$^{*\dagger}$ & 0.08 K -- 0.20 K \\
 Space Target Temperature & $< 0.01$ K& Spacecraft Temperature Contamination$^{*\dagger}$ & 0.01 K -- 0.05 K \\
 Space Target Emissivity & $< 0.01$ K& Radiometric Nonlinearity$^*$ & 0.01 K -- 0.08 K \\
 Reflected Space Target Temperature & 0.003 K& Antenna/Reflector Emissivity$^*$ & 0.01 K -- 0.02 K \\
 Radiometric Nonlinearity & $< 0.04$ K& Cosmic Microwave Background & 0.0006 K \\
 \hline \hline
\end{tabularx} 
\vspace{0.5ex} 
\hspace{0.5em}

\caption{ $*$Estimated based on radiometric contributions to calibration.\\
$\dagger$These uncertainties are correlated and thus cannot be considered independent.\\
Contributions to the radiometric uncertainty for the JPSS CrIS and JPSS ATMS instruments as proxy's for typical on-orbit Infrared and Microwave Sensors. The temperature knowledge of the internal calibration target (ICT) is a leading contributor to the total radiometric uncertainty.}
\label{tab:cris_atms}
\end{table*}

\section{Design Concepts}

\begin{figure}[b!]
    \centering
    \includegraphics[width=\linewidth]{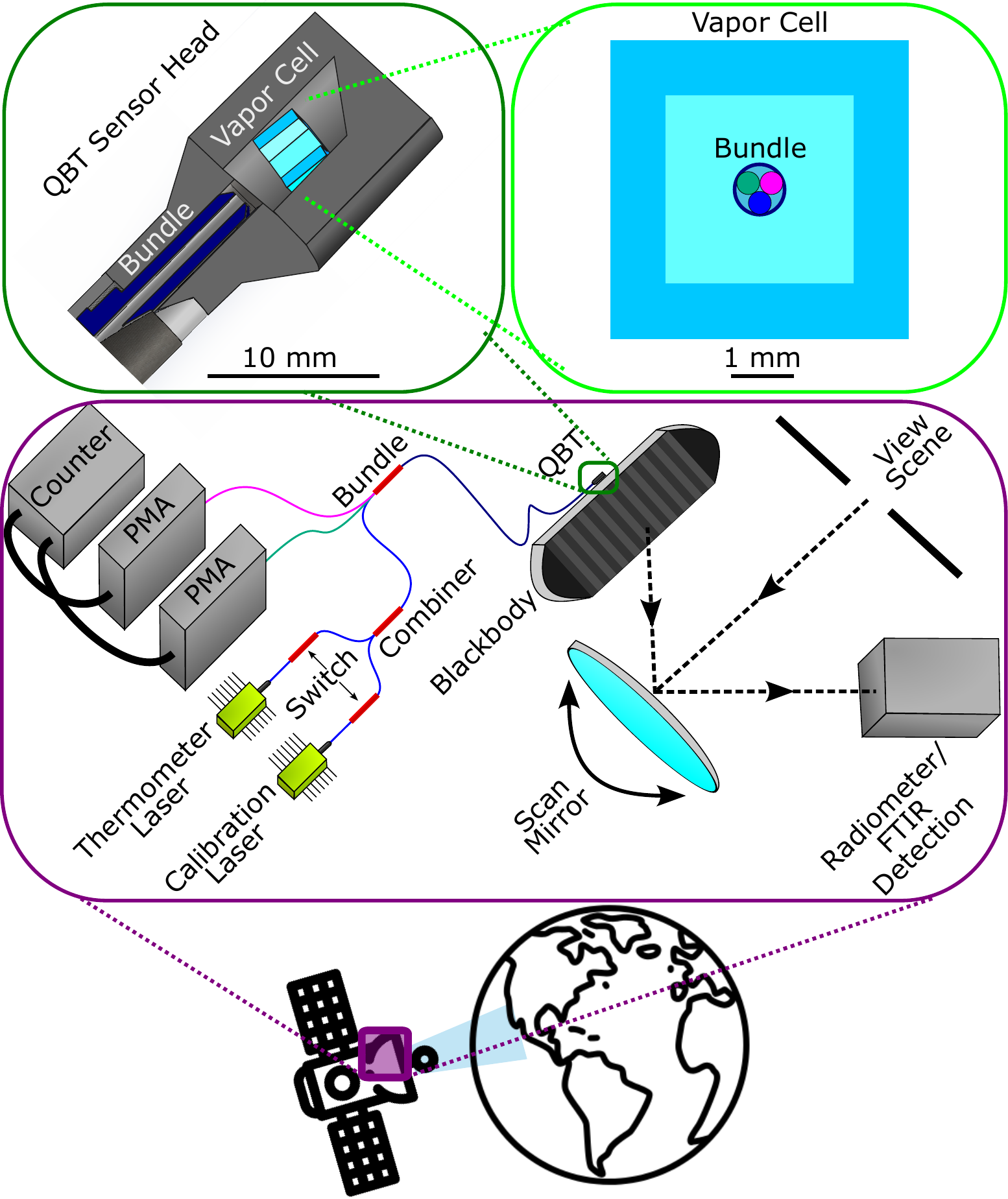}
    \caption{Schematic of a quantum blackbody thermometer (QBT) measuring an  on board radiometric calibration source blackbody. The QBT is based upon a microfabricated alkali vapor cell housed in a fiber-bundle-coupled sensor head.  Fiber switches select between exciting the alkali atoms with the thermometer and calibration lasers, and fluorescence photons are detected by photomultiplier assemblies (PMAs). Earth and satellite symbols adapted from icons by Lusi Astianah and JK Lim via The Noun Project.}
    \label{fig:schematic}
\end{figure}

Shown schematically in Fig.\,\ref{fig:schematic}, the QBT is centered on a microfabricated atomic vapor cell \cite{Liew2004,Kitching2016} in a fiberized sensor package.  
A bundle of three fibers interfaces with the vapor cell.  The first fiber delivers excitation light from alternately one of two lasers: the `thermometer laser' and the `calibration laser' (described in more detail in Section \ref{sec:iii}).  The other two fibers collect atomic fluorescence and deliver it to photon counting detectors, such as photomultiplier assemblies (PMAs). Fluorescence at wavelengths corresponding to particular atomic transitions are isolated using bandpass interference filters.  The signals from the PMAs are digitized using a fast multi-channel scalar or other counting device.

The response of a remote sensing instrument, such as a radiometer or FTIR detector, is periodically recalibrated against the OBRCS by adjusting the optical path using a scan mirror.  The QBT is mounted to the OBRCS in conjunction with other reference thermometers, such as PRTs.  
The PRTs can achieve  precision $\delta T/T = 3\times 10^{-7}$ on second time scales \cite{Strouse2007}, while the QBT could provide accuracy at the $1\times 10^{-4}$ level on minute time scales. 

This approach offers several additional technical advantages. First, the form factor of the QBT vapor cell (typically $3\times3\times3$\,mm$^3$) sensor with fiber readout is comparable to that of a typical PRT ($\sim 4$\,mm diameter, 50 mm long capsule) with four-wire readout, thus enabling drop-in replacement of PRTs with QBTs.  
Additionally, the QBT form factor is compatible with a commercial dry well calibrator, allowing for laboratory performance characterization.  Finally, the small thermal mass of the microfabricated cell minimizes temperature differences between the QBT and the thermal bath which it is sensing.

\section{Description of Proposed Technology}\label{sec:iii}

\begin{figure}[b!]
    \centering
    \includegraphics[width=\linewidth]{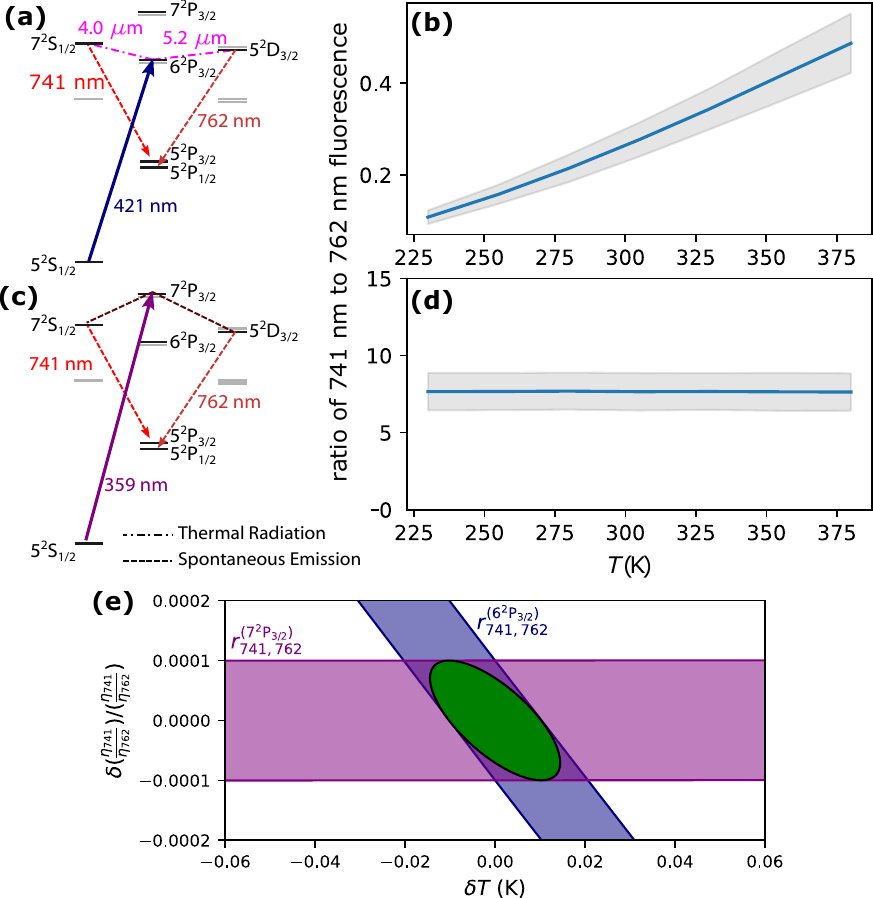}
    \caption{Proposed self-calibrating QBT sensing scheme.  a) Temperature sensing.  
    Rb atoms are laser-excited to the 6$^2$P$_{3/2}$ state by the 421\,nm thermometer laser.  
    TR stimulates transitions to 7$^2$S$_{1/2}$ and 5$^2$D$_{3/2}$ at different temperature-dependent rates.
    Relative populations of 7$^2$S$_{1/2}$ and 5$^2$D$_{3/2}$ are determined from the 741 nm and 762 nm fluorescence, respectively.
    b) Line shows the calculated ratio of 741 nm to 762 nm fluorescence when laser-exciting Rb with the thermometer laser.  
     Grey bands are 68\% confidence intervals extracted from Monte Carlo simulations which include theoretical uncertainty in the Rb TDMEs \cite{UDportal}.
    c) Detector calibration.  Rb atoms are laser-excited to the 7$^2$P$_{3/2}$ state by the 359\,nm calibration laser. 
    In this case, populations of 7$^2$S$_{1/2}$ and 5$^2$D$_{3/2}$ are due to spontaneous decay from 7$^2$P$_{3/2}$. 
    d) Thus, the calculated ratio of 741 nm to 762 nm fluorescence when exciting with the calibration laser is essentially temperature-independent, and provides sensitivity to the ratio of total detection efficiency at 741 nm to 762 nm. e) Example $1\sigma$ uncertainty plot constraining $T$ and $\eta_{741}/\eta_{762}$ (green) for the case of $r^{(6^2\rm{P_{3/2}})}_{741,762}$ and $r^{(7^2\rm{P_{3/2}})}_{741,762}$ each measured to a relative precision of $10^{-4}$ (shaded bands) at $T = 300$\,K .}
    \label{fig:ratios}
\end{figure}

Our method for measuring temperature is summarized in Fig.\,\ref{fig:ratios}. The method is based on the proposal and demonstration reported in Ref. \cite{LaMantia2025}, wherein temperature was determined by optical fluorescence ratios of laser-excited rubidium atoms in steady state.
Double excitation by laser light and thermal radiation (TR) induces fluorescence at multiple wavelengths at calculable, temperature-dependent rates.
By optically exciting different atomic transitions while monitoring the same fluorescence ratio, the known atomic transition dipole matrix elements (TDMEs) can be used to calibrate ratios of fluorescence detection efficiencies; that is, the atom provides the device an internal self-calibration  with long-term stability.

The QBT consists of an atomic vapor cell which acts as a contact thermometer, coming into thermal equilibrium with its environment. The vapor cell glass is chosen to be transparent for ultraviolet and visible light, but opaque for longer wavelengths. Thus, the vapor cell TR is well approximated by an ideal blackbody radiation (BBR) spectrum for infrared wavelengths. The TR emitted from the vapor cell glass drives transitions between excited atomic states at multiple infrared wavelengths. Temperature is then measured by monitoring optical fluorescence induced by one or more TR-stimulated atomic transitions.

We model the Rb atomic system as a function of radiometric temperature by solving a system of rate equations \cite{Schlossberger2024, Norrgard2021,LaMantia2025,Eckel2026} to determine the steady-state populations.  
The rate equation model includes spontaneous decay rates $\Gamma_{ij}$ and BBR-stimulated transition rates $\Omega_{ij}$ between states $\ket{i}$ and $\ket{j}$.   The model includes TDMEs between a total of 59 states of Rb (up to principal quantum numbers $n =$ 13, 13, 12, and 10  for $n^2$S$_{1/2}$, $n^2$P$_{1/2,\,3/2}$, $n^2$D$_{3/2,\, 5/2}$ and $n^2$F$_{5/2,\, 7/2}$ states, respectively) \cite{UDportal}.  The model further includes one or more laser excitation rates $R_{ij}$.
The steady state populations are determined by singular value decomposition of $M = \Gamma + \Omega + R $.

Figure \ref{fig:ratios} demonstrates one of several possible temperature sensing modalities. In Fig.\,\ref{fig:ratios}a, 421 nm light from a  thermometer laser excites Rb atoms to the 6$^2$P$_{3/2}$ state. 
TR stimulates transitions from 6$^2$P$_{3/2}$ to 7$^2$S$_{1/2}$ and to 5$^2$D$_{3/2}$ at different, temperature-dependent rates. 
The populations of the 7$^2$S$_{1/2}$ and 5$^2$D$_{3/2}$ states are determined from fluorescence measurements at 741 nm and 762 nm, respectively.
Fig. \ref{fig:ratios}b shows the value of $r^{(6^2\rm{P_{3/2}})}_{741,762}$, the 741 nm to 762 nm fluorescence ratio when laser-exciting to $6^2\rm{P_{3/2}}$, as a function of temperature $T$ in the full 59-state model.
In a simplified six-state model ignoring fine structure (5S, 5P, 6P, 5D, 7S, and 7P) and assuming $\Gamma_{ij} \gg \Omega_{ij}$ for any allowed spontaneous emission, we may solve analytically for the steady-state fluorescence signal ratio \cite{Eckel2026}
\begin{equation}\label{eq:simple thermometer}
    r^{(6\rm{P})}_{741,762} = \frac{\eta_{741}}{\eta_{762}} \frac{\Gamma_{5{\rm D}}}{\Gamma_{7{\rm S}}} \frac{\Gamma_{7{\rm S},5 {\rm P}}}{\Gamma_{5{\rm D},5{\rm P}}}\frac{ \Gamma_{7{\rm S},6{\rm P}}}{\Gamma_{5{\rm D},6{\rm P}} } \frac{(e^{hc/\lambda_{6{\rm P},5{\rm D}}kT}-1)}{(e^{hc/\lambda_{6{\rm P},7{\rm S}}kT}-1)}\,,
\end{equation}
where $\eta_\lambda$ is the total detection efficiency at wavelength $\lambda$ in nanometers, $\Gamma_i =\sum_j \Gamma_{ij}$, $h$ is the Planck constant, $k$ is the Boltzmann constant, $c$ is the speed of light, $\lambda_{6{\rm P},7{\rm S}} \approx 4.0\,\mu$m, and \mbox{$\lambda_{6{\rm P},5{\rm D}} \approx 5.2\,\mu$m}.

Apart from the detector efficiency ratio $\eta_{741}/\eta_{762}$, the temperature $T$ may be determined from a measurement of $r^{(6^2\rm{P_{3/2}})}_{741,762}$, fundamental constants, and atomic TDMEs.
Thus, $\eta_{741}/\eta_{762}$ must somehow be determined in order to assign a temperature value from a measured $r^{(6^2\rm{P_{3/2}})}_{741,762}$.
One method our group is currently exploring is to calibrate QBT fluorescence ratio $r^{(6^2\rm{P_{3/2}})}_{741,762}$ as a function of temperature by referencing to a high-accuracy transfer standard, such as a calibrated PRT. While subject to potential long-term drifts and shifts like any calibrated sensor, this method is simple in that it requires a single thermometer laser.

Alternately, $\eta_{741}/\eta_{762}$ may be determined by  exciting to another atomic state, such as the 7$^2$P$_{3/2}$ state (Fig.\,\ref{fig:ratios}c) using a calibration laser.  
In this case, the 741 nm to 762 nm fluorescence ratio is essentially temperature-independent (Fig.\,\ref{fig:ratios}d) and thus the ratio of total detection efficiencies $\eta_{741}/\eta_{762}$ may be determined entirely from the intrinsically stable atomic TDMEs with little regard to the temperature at which the calibration is performed. 
The measured $\eta_{741}/\eta_{762}$ then self-calibrates the temperature-dependent ratio in Fig.\,\ref{fig:ratios}a,b.
Returning to the simplified six-state model, the fluorescence signal ratio is 
\begin{equation}\label{eq:simple cal}
    r^{(7\rm{P})}_{741,762} = \frac{\eta_{741}}{\eta_{762}} \frac{\Gamma_{5{\rm D}}}{\Gamma_{7{\rm S}}} \frac{\Gamma_{7{\rm S},5 {\rm P}}}{\Gamma_{5{\rm D},5{\rm P}}}\frac{ \Gamma_{7{\rm P},7{\rm S}}}{\Gamma_{7{\rm P},5{\rm D}} }\,,
\end{equation}
so that Eq.\,\eqref{eq:simple thermometer} may be rewritten as
\begin{equation}\label{eq: calibrated}
    r^{(6\rm{P})}_{741,762} = r^{(7\rm{P})}_{741,762}  \frac{\Gamma_{7{\rm S},6 {\rm P}}}{\Gamma_{5{\rm D},6{\rm P}}}\frac{ \Gamma_{7{\rm P},5{\rm D}}}{\Gamma_{7{\rm {D}},7{\rm S}} } \frac{(e^{hc/\lambda_{6{\rm P},5{\rm D}}kT}-1)}{(e^{hc/\lambda_{6{\rm P},7{\rm S}}kT}-1)}\,.
\end{equation}

The simplified six-state model is useful insofar as it identifies the key aspects of the temperature measurement:
excitation by either the calibration laser or the thermometer laser provides sensitivity to the total detection efficiency ratio  $\eta_{741}/\eta_{762}$, while temperature sensitivity is achieved primarily by excitation with the thermometer laser.
A more complete description of the QBT is that $r^{(6^2\rm{P_{3/2}})}_{741,762}$ and $r^{(7^2\rm{P_{3/2}})}_{741,762}$ are measured sequentially and with sufficient precision to constrain both $T$ and $\eta_{741}/\eta_{762}$.  For example, Fig.\,\ref{fig:ratios}e highlights the case of measuring $r^{(6^2\rm{P_{3/2}})}_{741,762}$ and $r^{(7^2\rm{P_{3/2}})}_{741,762}$ each to a relative precision of $10^{-4}$ at $T = 300$\,K yields relative precision of $10^{-4}$  on $\eta_{741}/\eta_{762}$ and $5\times 10^{-5}$ on $T$.



\section{Projected QBT Performance}
In order to validate the QBT performance metrics, it should be studied and calibrated in a known temperature environment.
Temperature uniformity of the calibration environment is likely to limit the device calibration accuracy.  In Table \ref{tab:placeholder} we compare expected QBT performance to that of typical calibrated PRTs and to the proof-of-concept demonstration of Ref.\,\cite{LaMantia2025}.
We assume 20\,mK uniformity [$u(T)/T = 7\times 10^{-5}$ at $T= 300$\,K], typical of commercial dry well calibrators, limits the calibration accuracy.   We note that fluid bath calibrators are often able to provide temperature uniformity of several mK.

\begin{table}[b!]
    \centering
    \begin{tabular}[width= \columnwidth]{llll}
    \hline\hline
    Specification & PRT \cite{Mason1996,Smith2020, Smith2021,Tobin2013,Latvakoski2011,Strouse2007}& Ref.\,\cite{LaMantia2025} & QBT Target \\
  \hline
    Drift $\frac{\Delta T}{T} \frac{1}{\rm{year}}$    & $3 \times 10^{-5}$ & --- & 0$^\dagger$ \\
   Instantaneous shift $\frac{\Delta T}{T}$& $< 1\times10^{-3}$ & ---& $1 \times 10^{-4}$$^\dagger$  \vspace{4pt}\\
    
                        & Essentially\\
    Sensitivity $\frac{\delta T}{T}\frac{1}{\sqrt{t/\text{s}}}$ &  instantaneously   & $1\times10^{-3}$ &  $3\times10^{-4}$\\
                        &noise-limited \vspace{4pt}\\

    Precision $\frac{\delta T}{T}$ & $<1 \times 10^{-6}$  & $4 \times 10^{-4}$ & $1\times10^{-5}$\vspace{4pt}\\

Accuracy $\frac{u(T)}{T}$ \\
    
    \quad\quad Calibrated       & $<1 \times 10^{-6}$  & $1.2 \times 10^{-2}$ & $7 \times 10^{-5}$\\
    \quad\quad Self-calibrated & ---                    & ---               & $1 \times 10^{-4}$\\
    \quad\quad Typical on-orbit$^*$  & $1\times10^{-3}$ & ---& $1 \times 10^{-4}$ \\
    \hline\hline
    \end{tabular}
    
    \caption{$^\dagger$ With self-calibration\\
    *Assuming a nominally 300\,K OBRCS\\
    Comparison of proposed QBT performance to typical on-orbit PRTs and to the demonstration of Ref. \cite{LaMantia2025}.}
    \label{tab:placeholder}
\end{table}

The accuracy of the self-calibration is currently limited to roughly 15\,\% by combined uncertainty in the TDMEs of Rb \cite{LaMantia2025,UDportal}.  Shown as the gray bands in Fig.\,\ref{fig:ratios}, the $k=1$ model uncertainties in the fluorescence ratios are determined by a Monte Carlo simulation where each TDME is varied about its mean value using a normal distribution with a width equal to the theoretical uncertainty \cite{UDportal}.
The uncertainty is likely overestimated by assuming the TDME uncertainties are uncorrelated; for example, the $n\,^2$P$_{3/2}$ to $n^\prime\,^2$D$_{3/2}$ and $n\,^2$P$_{3/2}$ to $n^\prime\,^2$D$_{5/2}$ TDMEs should be almost perfectly correlated, with their ratio determined from Clebsch-Gordan coefficients.
Regardless of the current estimated model uncertainty, the QBT response to excitation by the calibration laser and thermometer laser may be calibrated to a  primary or transfer temperature standard.
To the extent that other systematic errors can be mitigated, this one-time calibration of the atomic response to the thermometer and calibration lasers may be transferred to any future implementation of the QBT.

In order to achieve a target temperature sensitivity of $\delta T/T = 3\times 10^{-4}/\sqrt{t/\text{s}}$, each fluorescence wavelength must be detected with a photon count rate of roughly $10^7$\,s$^{-1}$, typical of fast multi-channel counters inclusive of  a margin of safety for potential pulse pileup effects. 
Assuming Poissonian statistics, the targeted temperature precision $\delta T/T = 1\times 10^{-5}$ will require an averaging time of 900 s. 
We note that this level of temperature precision is only necessary for investigating sources of systematic error, as it greatly exceeds our expected transfer standard-calibrated accuracy ($7\times 10^{-5}$) and self-calibrated accuracy ($1\times 10^{-4}$).  The expected averaging time to achieve a precision equal to these target accuracies are 18 s and 9 s, respectively.  In order to usefully contribute to long-term performance, the expected accuracy of the QBT should exceed typical intersensor discrepancies, up to 160 mK or $5\times 10^{-4}$ at $T = 300$\,K  
\cite{Loveless2023}.

The QBT is expected to operate over a temperature range of roughly 250\,K to 400\,K.  The lower temperature limit is set by the reduced signal due to reduced Rb vapor pressure at lower temperature.  This could possibly be mitigated by increased thermometer laser power or by light-induced atomic desorption \cite{Gozzini1993}, at the expense of increased self-heating effects.  Cs has roughly five times higher vapor pressure than Rb at a given temperature. A fluorescence intensity ratio temperature sensing scheme with Cs could also be developed to optimize for a lower temperature range.  
The QBT is expected to be limited at high temperatures by state-changing Rb-Rb collisions which induce additional excitation and relaxation processes. Although the Rb vapor pressure increases with increasing temperature, Ref. \cite{LaMantia2025} found Rb-Rb collisions were not statistically significant at the $\delta T/ T = 4\times 10^{-4}$ level for temperatures up to 343\,K. However, inclusion of state-changing collisions in the rate-equation model may be necessary to accurately calculate fluorescence ratios at higher temperatures. The lower vapor pressure of K compared to Rb could make a fluorescence intensity ratio temperature sensing scheme with K more appropriate when optimizing for a higher temperature range. For yet higher temperatures, collisions were found to be negligible in microfabricated Sr vapor cells at temperatures up to 573\,K \cite{Pate2023}.

\begin{figure}[b!]
    \centering
    \includegraphics[width=\linewidth]{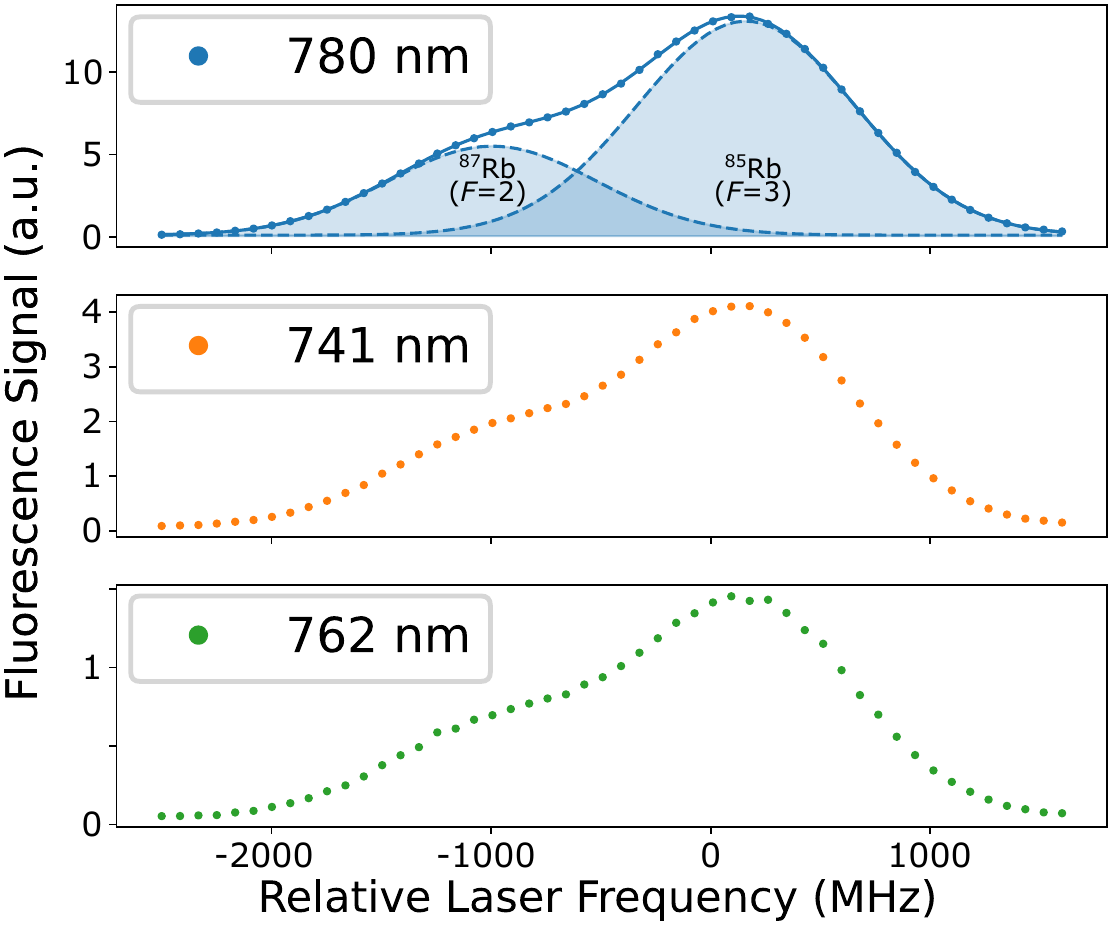}
    \caption{Measured prototype QBT fluorescence (dots) as a function of laser frequency as the laser is scanned over the upper ground hyperfine manifold of the 5$^2$S$_{1/2} \rightarrow 7^2$P$_{3/2}$ transition for $^{87}$Rb and $^{85}$Rb.    Fluorescence at 741\,nm and 762\,nm demonstrate a high signal-to-noise ratio for the self-calibration scheme in Fig.\,\ref{fig:ratios}.  Strong fluorescence is also detected on the $5^2$P$_{3/2} \rightarrow 5^2$S$_{1/2}$ decay at 780\,nm.  The solid line is a fit of two Gaussian peaks (one for each isotope) plus constant offset to the 780\,nm fluorescence, with excellent agreement to the data. The shaded dashed lines show the constituent single Gaussian peaks.  }
    \label{fig:spectra}
\end{figure}

As a preliminary test, we  constructed a prototype QBT shown in the upper left of Fig.\,\ref{fig:schematic}.
Rubidium atoms were excited to the 7$^2$P$_{3/2}$ transition by roughly $50\,\mu$W of power from the calibration laser.
Fluorescence from different atomic transitions was selected using bandpass interference filters with 10 nm full width at half maximum prior to the PMA.
Figure \ref{fig:spectra} shows the detected fluorescence signals as a function of calibration laser frequency with the QBT held in a dry well calibrator at $T=313$\,K.  
In addition to the 741\,nm and 762\,nm fluorescence relevant to the sensing scheme outlined in Fig.\,\ref{fig:ratios}, we also detect 780\,nm fluorescence from the 5$^2$P$_{3/2} \rightarrow 5^2$S$_{1/2}$ transition.  The peak 741\,nm and 762\,nm fluorescence count rate is sufficient to determine $\eta_{741}/\eta_{762}$ to roughly $1\times 10^{-3}$ in 1\,s. All else equal, equivalent count rates should be possible with $\sim 10$\,mW of thermometer laser power, which would determine $T$ to roughly $5\times 10^{-4}$ in 1\,s  This is only a factor of 2 worse than the targeted temperature sensitivity of Table \ref{tab:placeholder}, and we expect substantial improvements are possible through a combination of improved fiber collection efficiency and reduced background.



\section{Conclusion}

Here we have proposed use of a QBT for on-orbit remote sensing absolute temperature reference. The QBT is self-calibrating, fast, and does not drift over time. We have presented a roadmap to validate these features in a form factor suitable for on-orbit sensing, and demonstrated a prototype QBT with sensitive, fully fiber-coupled excitation and fluorescence readout. 

Temperature is at the core of nearly all science observations.  For remote sensing activities, the temperature of instruments such as cameras, motors, batteries, mirrors, and more must be continuously monitored, and often controlled, to ensure their intended operation and longevity.   
Improvement of thermometers at the component-level for spaceborne measurements will enhance the long-term accuracy of all of these instruments.  

Among many possible use cases, our targeted QBT performance should substantially reduce the uncertainty in the temperature of OBRCSs, one of the leading uncertainties in on-orbit sensors' error budgets. 
Accurate weather prediction and modeling relies upon accurately calibrated remote sensing thermometers and comparison to accurate historical data.  
Planetary missions take several years to reach their observation targets; our QBT will ensure that temperature measurements on future planetary missions are as accurate at launch as they are when investigating the history and potential habitability of our solar system.  

\section*{Acknowledgments}
We thank Dixith Manchaiah, Sai Naga Manoj Paladugu, and Weston Tew for helpful discussions, and Eric Shirley and Matthew Hummon  for a thorough reading of the manuscript.   This work was supported in part by NIST. The authors acknowledge the use of icons created by Lusi Astianah and JK Lim from The Noun Project, used under CC-BY 3.0.  The views, opinions and/or findings expressed are those of the authors and should not be interpreted as representing the official views or policies of the the U.S. Government. A contribution of the U.S. government, this work is not subject to copyright in the United States.





\bibliographystyle{IEEEtran}
\bibliography{reference}
\end{document}